\documentclass[12pt]{article}

\usepackage{scicite}
\usepackage{times}

\usepackage[T1]{fontenc} 
\usepackage[utf8]{inputenc} 
\usepackage{float} 
\usepackage{graphicx} 

\newenvironment{sciabstract}{%
\begin{quote} \bf}
{\end{quote}}

\title{An extreme test case for planet formation: a close-in Neptune orbiting an ultracool star}

\author
{Guðmundur Stefánsson,$^{1\ast}$, Suvrath Mahadevan$^{2,3,4}$, Yamila Miguel$^{5,6}$,\\
Paul Robertson$^{7}$, Megan Delamer$^{2,3}$, Shubham Kanodia$^{8}$, Caleb Cañas$^{9}$\\
Joshua Winn$^{1}$, Joe Ninan$^{10}$, Ryan Terrien$^{11}$, Rae Holcomb$^{7}$, Eric Ford$^{2,3,12,13}$\\
Brianna Zawadzki$^{2,3}$, Brendan P. Bowler$^{14}$, Chad Bender$^{15}$, William Cochran$^{16}$\\
Scott Diddams$^{17,18,19}$, Michael Endl$^{16,14}$, Connor Fredrick$^{18,19}$,\\
Samuel Halverson$^{20}$, Fred Hearty$^{2,3}$, Gary J. Hill$^{21,14}$, Andrea Lin$^{2,3}$\\
Andrew Metcalf$^{22,18,19}$, Andrew Monson$^{15}$, Lawrence Ramsey$^{2,3}$, Arpita Roy$^{23,24}$\\
Christian Schwab$^{25}$, Jason Wright$^{2,3,26}$, Gregory Zeimann$^{27}$\\
\normalsize{$^{1}$Princeton University, Department of Astrophysical Sciences, USA}\\
\normalsize{$^{2}$Department of Astronomy \& Astrophysics, The Pennsylvania State University, USA}\\
\normalsize{$^{3}$Center for Exoplanets and Habitable Worlds, The Pennsylvania State University, USA}\\
\normalsize{$^{4}$ETH Zurich, Institute for Particle Physics \& Astrophysics, Switzerland}\\
\normalsize{$^{5}$Leiden Observatory, University of Leiden, The Netherlands}\\
\normalsize{$^{6}$SRON Netherlands Institute for Space Research, The Netherlands}\\
\normalsize{$^{7}$Department of Physics \& Astronomy, University of California, Irvine, USA}\\
\normalsize{$^{8}$Earth and Planets Laboratory, Carnegie Institution for Science, USA}\\
\normalsize{$^{9}$NASA Goddard Space Flight Center, USA}\\
\normalsize{$^{10}$Department of Astronomy \& Astrophysics, Tata Institute of Fundamental Research, India}\\
\normalsize{$^{11}$Carleton College, USA}\\
\normalsize{$^{12}$Center for Astrostatistics, The Pennsylvania State University, USA}\\
\normalsize{$^{13}$Institute for Computational \& Data Sciences, The Pennsylvania State University, USA}\\
\normalsize{$^{14}$Department of Astronomy, The University of Texas at Austin, Austin, TX 78712, USA}\\
\normalsize{$^{15}$Steward Observatory, The University of Arizona, USA}\\
\normalsize{$^{16}$Center for Planetary Systems Habitability and McDonald Observatory, UT Austin, USA}\\
\normalsize{$^{17}$Electrical, Computer and Energy Engineering, University of Colorado, USA }\\
\normalsize{$^{18}$National Institute of Standards \& Technology,USA}\\
\normalsize{$^{19}$Department of Physics, University of Colorado, USA}\\
\normalsize{$^{20}$Jet Propulsion Laboratory, California Institute of Technology, USA}\\
\normalsize{$^{21}$McDonald Observatory, UT Austin, USA}\\
\normalsize{$^{22}$Space Vehicles Directorate, Air Force Research Laboratory, USA}\\
\normalsize{$^{23}$Space Telescope Science Institute, USA}\\
\normalsize{$^{24}$Department of Physics and Astronomy, Johns Hopkins University, USA}\\
\normalsize{$^{25}$School of Mathematical and Physical Sciences, Macquarie University, Australia}\\
\normalsize{$^{26}$Penn State Extraterrestrial Intelligence Center, The Pennsylvania State University, USA}\\
\normalsize{$^{27}$Hobby-Eberly Telescope, University of Texas, Austin, Austin, TX, 78712, USA}\\
\normalsize{$^\ast$To whom correspondence should be addressed; E-mail: gstefansson [at] princeton [dot] edu} 
}

\date{Submitted Manuscript to Science on October 17, 2022. In review.}

\begin{document} 
\baselineskip24pt
\maketitle

\begin{sciabstract} 
In current theories of planet formation, close-orbiting planets as massive as Neptune are expected to be very rare around low-mass stars. We report the discovery of a Neptune-mass planet orbiting the `ultracool' star LHS 3154, which is nine times less massive than the Sun. The planet's orbital period is 3.7 days and its minimum mass is 13.2 Earth masses, giving it the largest known planet-to-star mass ratio among short-period planets ($<$\,100 days) orbiting ultracool stars. Both the core accretion and gravitational instability theories for planet formation struggle to account for this system. In the core-accretion scenario, in particular, the dust mass of the protoplanetary disk would need to be an order of magnitude higher than typically seen in protoplanetary disk observations of ultracool stars.
\end{sciabstract}

\bigskip

\noindent {\bf One-Sentence Summary:} \\ 
A Neptune-mass planet found orbiting the ultracool star LHS 3154 provides an extreme test case for planet formation theories.

\noindent {\bf Main Text:} 

In the spectral classification of stars, the lowest mass stars---the `M dwarfs'---are the most common stars in the solar neighborhood and throughout the Galaxy \cite{henry2006,winters2015}. Gas giant planets are much rarer around M dwarfs than the more massive F, G, and K dwarfs \cite{endl2006}, and planets orbiting M dwarfs tend to be less massive than Neptune \cite{bonfils2013,sabotta2021}. However, the properties of planets around the least massive ($<0.25$ solar masses) and coolest M dwarfs---the `ultracool' dwarfs---are still poorly known. This is because ultracool dwarfs are faint and emit most of their radiation at infrared wavelengths, where planet-finding technologies are not as well developed as at optical wavelengths.

Two emblematic planetary systems around ultracool dwarfs are TRAPPIST-1 \cite{gillon2017} and Teegarden's star \cite{zechmeister2019}, which both feature compact systems of small and potentially rocky planets. The formation of such systems is compatible with the theory of core accretion \cite{alibert2017,miguel2020,zawadzki2021,burn2021}, where the outcome of planet formation depends strongly on the assumed total mass of small solid particles (`dust') within the protoplanetary disk \cite{miguel2020}. Observations of protoplanetary disks show that the dust mass tends to be lower for low-mass stars \cite{ansdell2016,pascucci2016}, making it possible to form Earth-mass planets but not much more massive planets. However, the uncertainties in the models and the large observed dispersion in dust masses raise the intriguing possibility that planetesimal core accretion might, at least occasionally, allow a low-mass star to form close-orbiting planets with a masses exceeding $\sim$10 Earth masses.

Massive planet candidates have indeed been detected around a few ultracool stars, but in all cases the planets have very wide orbits. The clearest examples are GJ\,3512 b (minimum mass of $0.46 M_{\mathrm{jup}}$; $P=203$days) \cite{morales2019}, and TZ\,Ari\,b (minimum mass of $0.21 M_{\mathrm{jup}}$; $P=771$days) \cite{quirrenbach2022}. These gas giants are thought to have formed from a different mechanism than core accretion, such as gravitational instability within a massive gaseous outer disk, which is much more efficient at producing wide-orbiting planets than close-orbiting planets \cite{mercer2020}. To this point, close-in gas giants or Neptune-mass planets have not been discovered around ultracool stars.

We announce the discovery of a 13.2 Earth-mass planet (minimum mass) orbiting the M6.5 dwarf LHS 3154, located 16\,pc away from the Sun. We observed LHS 3154 using the Habitable-zone Planet Finder (HPF) Spectrograph \cite{mahadevan2012,mahadevan2014}, a near-infrared spectrograph on the 10m Hobby-Eberly Telescope (HET) at McDonald Observatory in Texas \cite{ramsey1998,hill2021}. We observed LHS 3154 as part of the HPF Survey to detect and characterize the planet population around nearby ultracool stars. Observations between January 23 2020 and April 13 2022 revealed a periodic Doppler shift with a period of 3.7 days and a false-alarm probability much less than $0.1$\%. Figure 1 shows the radial velocity (RV) data. The RV residuals reveal no clear evidence of another planet in the system.

Table 1 summarizes the properties of the LHS 3154 system. We estimate the stellar mass and radius to be $0.1118\pm0.0027M_\odot$ and $0.1405\pm0.0038R_\odot$, respectively, based on the relationships of \cite{mann2015} and \cite{mann2019}. We determined the stellar effective temperature ($T_{\mathrm{eff}}= 2861 \pm 77$K) using the \texttt{HPF-SpecMatch} \cite{stefansson2020a} code, which compares a given HPF spectrum to a library of other well-characterized spectra with known properties. The metallicity of the star, which is difficult to precisely constrain for ultracool stars, was found to be consistent with solar. The low derived eccentricity of LHS 3154b of $e=0.076_{-0.047}^{+0.057}$ is consistent with a circular orbit.

Intrinsic variations of a star's atmosphere---known as `stellar activity'---can produce apparent radial-velocity shifts that may masquerade as a planetary signal \cite{robertson2014}. We used several metrics to assess the stellar activity of LHS 3154. We found no correlation between the RV signal and activity indicators in the HPF spectra (see Supplementary Materials). Furthermore, data from the Low-Resolution Spectrograph (LRS2) \cite{chonis2016} on the HET (resolution $R=2,500$) showed no evidence of H$\alpha$ emission. \cite{newton2017} showed that ultracool M dwarfs without detectable H$\alpha$ emission are slowly rotating; based on their results, we expect the rotation period of LHS 3154 to be $114\pm22$ days. Such slow rotation is consistent with the lack of discernible rotational broadening in the HPF spectra. Time-series photometry from the Transiting Exoplanet Survey Satellite \cite{ricker2015} shows no evidence for flaring, nor starspots. Time-series photometry over a longer timespan from the Zwicky Transient Facility (ZTF) \cite{masci2019} shows no evidence for variability on timescales near the 3.7-day period of the planetary signal, but does show evidence for variations on timescales between 90 and 140 days, consistent with the slow rotation of an inactive star. From these lines of evidence, we conclude that LHS 3154b is a slowly-rotating inactive star.

From the RVs, we found $m \sin i = 13.15_{-0.82}^{+0.84} M_\oplus$, where the orbital inclination, $i$, is unknown. To derive an upper limit on the planet's mass, we used astrometric information from the \textit{Gaia} spacecraft. A sufficiently massive companion would have caused detectable astrometric motion, manifested as `excess astrometric noise' \cite{Kervella2019}. For LHS 3154, the Gaia Data Release 3 reported an excess astrometric noise of 316 $\mu as$ with a significance of 27.7$\sigma$. However, the available data do not specify the timescale of the astrometric variability, making it impossible to determine whether it is from the 3.7-day planet or an additional long-period companion. By assuming all of the observed excess astrometric noise is due to the 3.7-day planet, and following the procedure of \cite{Penoyre2020} assuming a single-star solution, we obtained a $3\sigma$ upper mass limit of $32\mathrm{M_J}$. This rules out the possibility that the RV variations are due to a stellar binary companion. To have a mass above the brown dwarf limit ($\geq13M_J$), LHS 3154b would need an orbital inclination $i\leq0.2^\circ$. Although this has a very low a priori probability ($p\sim 10^{-6}$ for an isotropic distribution of inclination values), RV surveys are known to detect such face-on systems as planet candidates (e.g., \cite{wright2013}). Acknowledging this, but given the known low occurrence rate of brown dwarfs in short period orbits, coupled with the \textit{Gaia} constraint of $<32\mathrm{M_J}$ for any possible companion, we strongly prefer the planetary hypothesis.

Figure 2 compares the planet-to-star mass ratio of LHS 3154b to planets around other ultracool M-dwarfs with precise mass measurements. LHS 3154b is the only short-period Neptune-mass planet known to orbit an ultracool dwarf. Indeed, the mass ratio between the planet and the star ($3.5\times10^{-4}$) is the largest of all the known short-period ($<$100days) planets orbiting ultracool stars. With such a large mass ratio, LHS\,3154b presents an extreme test case for theories of planet formation around low-mass stars.

One way to form planets is within the core-accretion mechanism, where planets form from initial cores that accrete dust and gas. The expectations from core-accretion models where initial cores grow through the accretion of big planetesimals (see e.g., \cite{alibert2017,miguel2020,zawadzki2021,pan2022}) and core-accretion models supplemented with accretion from pebble-sized material (see e.g., \cite{liu2020,dash2020}) are that ultracool stars tend to form compact systems of terrestrial planets on short-period orbits. In particular, \cite{miguel2020} found that the maximum mass of planets formed by core accretion around low-mass stars is about $5 M_\oplus$. Thus, with a minimum mass of $13.2 M_\oplus$, LHS 3154b presents a challenge to this theory.

The outcomes of planet formation models depend sensitively on the assumed disk properties, including the total disk mass and gas-to-dust ratio. To gain further understanding of the disk properties required to form LHS 3154b, we performed detailed planet formation simulations for this system based on the core accretion scenario using the model of \cite{miguel2020}, modified by adding the possibility of gas accretion, following \cite{ida2004} (details in Supplement). 

Figure 3 summarizes the results. Figure 3a shows a simulation with $M_{\mathrm{disk}} = 0.01M_\star$, a typical assumption for the protoplanetary disk mass around a $0.1M_\odot$ star \cite{williams2011}, and $M_{\mathrm{dust}}/M_{\mathrm{gas}}=1\%$, typical of the interstellar medium \cite{bohlin1978}. Such simulations were found to be incapable of forming a close-orbiting planet as massive as LHS 3154b. To form LHS 3154b, we need an order-of-magnitude increase in the disk mass (Figure 3b), the dust-to-gas ratio (Figure 3c), or both (Figure 3d). Larger amounts of solid material enable the formation of more substantial planetary cores that can grow into more massive planets. 

Another possibility to form massive planets is through the gravitational instability mechanism, which has been invoked to form massive gas giant planets around low-mass stars. For example, \cite{morales2019} discussed the gravitational instability mechanism as a likely explanation for the widely-orbiting gas giant planet GJ 3512b (minimum mass of $0.46 M_{\mathrm{jup}}$; $P=203$days). However, LHS 3154b's minimum mass of $13.2 M_\oplus$ is significantly lower than the minimum mass planets formed from gravitational instability: simulations from \cite{morales2019} for a 0.1MSun star suggest minimum mass fragments of $\sim$60$M_\oplus$---about 5 times larger than the minimum mass of LHS 3154b. Although we can not decisively rule out the gravitational instability mechanism, if LHS 3154b indeed formed via gravitational instability, this would require even higher protoplanetary disk masses than we considered for the core-accretion scenario for the disks to be gravitationally unstable. As such, in order to form LHS 3154b, both formation mechanisms point to protoplanetary disks that are substantially more massive than typically seen in protoplanetary disk observations of ultracool stars.

One possibility to solve for the missing dust is that a significant fraction of the dust content of disks around low mass stars could have grown to centimeter sizes or more (e.g., \cite{williams2012}), and is therefore not detected with millimeter observations. Another possibility is that protoplanetary cores form at very early ages, well before 1 million years when disks are expected to be more massive.

The detection of LHS 3154b confirms for the first time that ultracool dwarfs can form massive close-in planets. Using the LHS 3154b detection among non-detections in 18 other ultracool stars surveyed as part of the HPF Survey, using the formalism of \cite{sabotta2021}, we calculate an initial occurrence estimate of $5_{-3}^{+8}\%$ of Neptune-mass planets (minimum masses from $10-100M_\oplus$) on short-period orbits ($1-10$days) around ultracool stars with $<0.25M_\odot$. This is consistent with the occurrence rate of $<10\%$ for $10-100M_\oplus$ minimum mass planets on short-period orbits reported by \cite{sabotta2021} for $<0.34 M_\odot$ stars. Although this occurrence estimate will be further refined as the HPF Survey is completed, this initial occurrence rate can be further tested with other exoplanet detection methods including the SPECULOOS survey \cite{delrez2022} which employs the transit technique to look for planets around ultracool stars.


\section*{Acknowledgments}

\textbf{Funding:} This work was partially supported by funding from the Center for Exoplanets and Habitable Worlds. The Center for Exoplanets and Habitable Worlds is supported by the Pennsylvania State University, the Eberly College of Science, and the Pennsylvania Space Grant Consortium. This work was supported by NASA Headquarters under the NASA Earth and Space Science Fellowship Program through grants 80NSSC18K1114. We acknowledge support from NSF grants AST-1006676, AST-1126413, AST-1310885, AST-1517592, AST-1310875, the NASA Astrobiology Institute (NAI; NNA09DA76A), and PSARC. We acknowledge support from the Heising-Simons Foundation via grant 2017-0494 and 2019-1177. G.S. acknowledges support provided by NASA through the NASA Hubble Fellowship grant HST-HF2-51519.001-A awarded by the Space Telescope Science Institute, which is operated by the Association of Universities for Research in Astronomy, Inc., for NASA, under contract NAS5-26555. Computations for this research were performed on the Pennsylvania State University’s Institute for Computational \& Data Sciences (ICDS). The research was carried out, in part, at the Jet Propulsion Laboratory, California Institute of Technology, under a contract with the National Aeronautics and Space Administration (80NM0018D0004). The Hobby-Eberly Telescope (HET) is a joint project of the University of Texas at Austin, the Pennsylvania State University, Ludwig-Maximilians-Universität München, and Georg-August-Universität Göttingen. The HET is named in honor of its principal benefactors, William P. Hobby and Robert E. Eberly. The Low Resolution Spectrograph 2 (LRS2) was developed and funded by the University of Texas at Austin McDonald Observatory and Department of Astronomy and by Pennsylvania State University. We thank the Leibniz-Institut für Astrophysik Potsdam (AIP) and the Institut für Astrophysik Göttingen (IAG) for their contributions to the construction of the integral field units.

\textbf{Author Contributions:} G.S. led the writing and interpretation of the result, including analyzing the RVs and host star properties. S.M. is PI of the HPF instrument, provided oversight, and assisted with analysis and interpretation. Y.M., in discussion with G.S. led the planet formation simulations. P.R. performed starspot simulations and aided in the detailed interpretation of the result. M.D. and G.S. performed the HPF RV sensitivity and occurrence analysis. C.C. performed the Gaia upper mass limit constraints. J. Winn provided expertise into the interpretation of the result and editing of the manuscript. G.Z., S.K., G.H. and G.S. reduced and interpreted the LRS2 spectra. B.Z. and E.F. contributed to the interpretation of planet formation. R.H. aided with the TESS photometric analysis. J.N., C.B., A.R., and R.T. extracted the HPF spectra and provided accurate wavelength calibration. S.D., A.Metcalf, and C.F. led the development of the HPF Laser Frequency Comb Calibrator. B.B. provided expertise into brown dwarf occurrence rates. W.C. and M.E. led observing programs contributing observing time. S.H., F.H., A.L., A.Monson, L.R., C.S., and J.Wright are part of the HPF team. All authors contributed to the interpretation of the result.

\textbf{Competing interests:} The authors declare no competing interests.

\textbf{Data and materials availability:} All time series data used (RVs, and activity indices) are available as machine-readable tables. The \texttt{juliet} code is available at https://github.com/nespinoza/juliet. The \texttt{HPF-SpecMatch} code is available at https://gummiks.github.io/hpfspecmatch/. The Panacea code is available at https://github.com/grzeimann/Panacea/.

\clearpage

\begin{figure}[t!]
\begin{center}
\includegraphics[width=\textwidth]{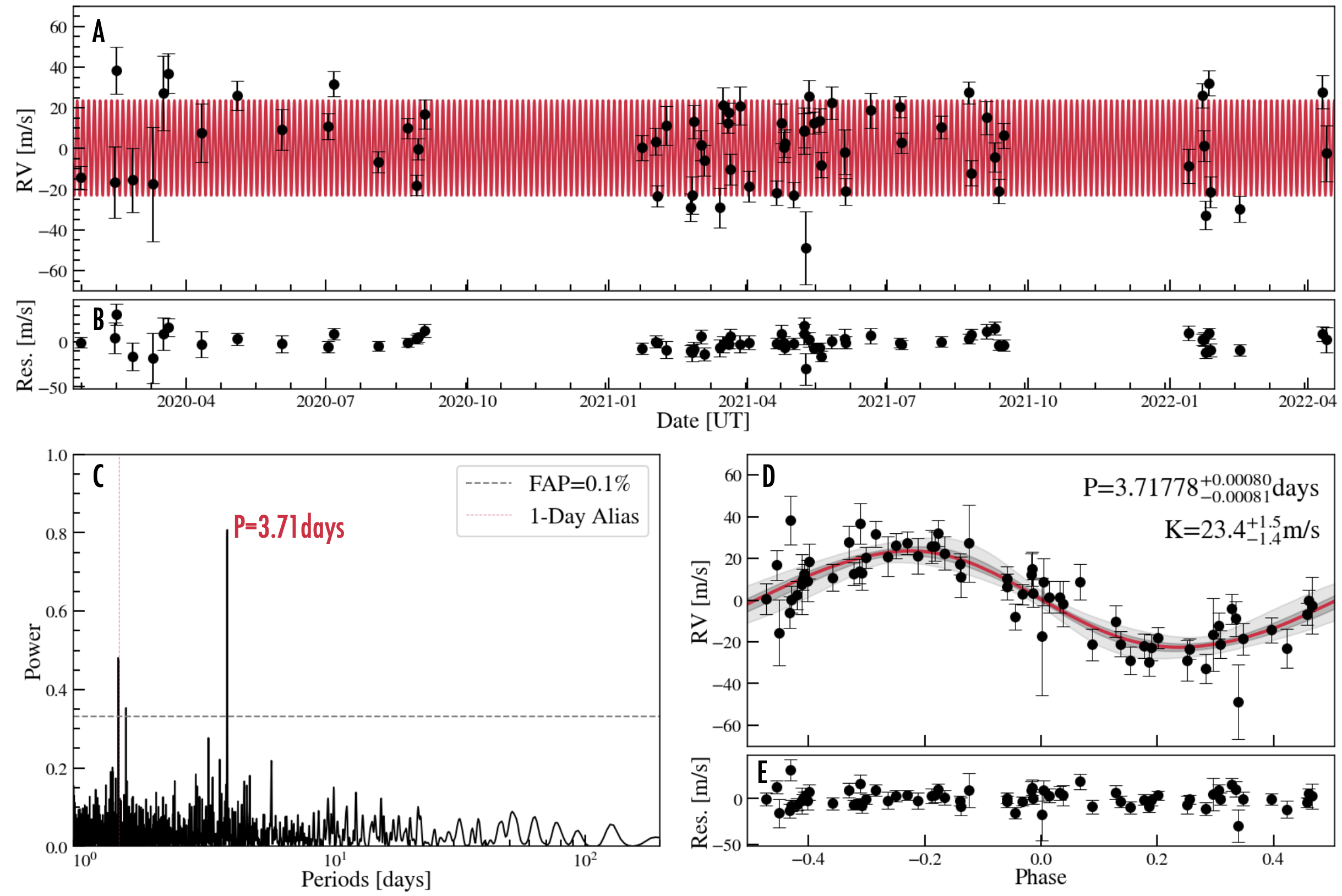}
\end{center}
\end{figure}
\noindent {\bf Fig~1.} HPF observations of LHS 3154b. \textbf{(A-B)} Radial velocity vs.\ time, and associated residuals. The orbital model is shown in red. \textbf{(C)} Periodogram, with a peak at 3.7 days. \textbf{(D-E)} Phase-folded radial velocities and associated residuals.

\clearpage

\begin{figure}[t!]
\begin{center}
\includegraphics[width=\textwidth]{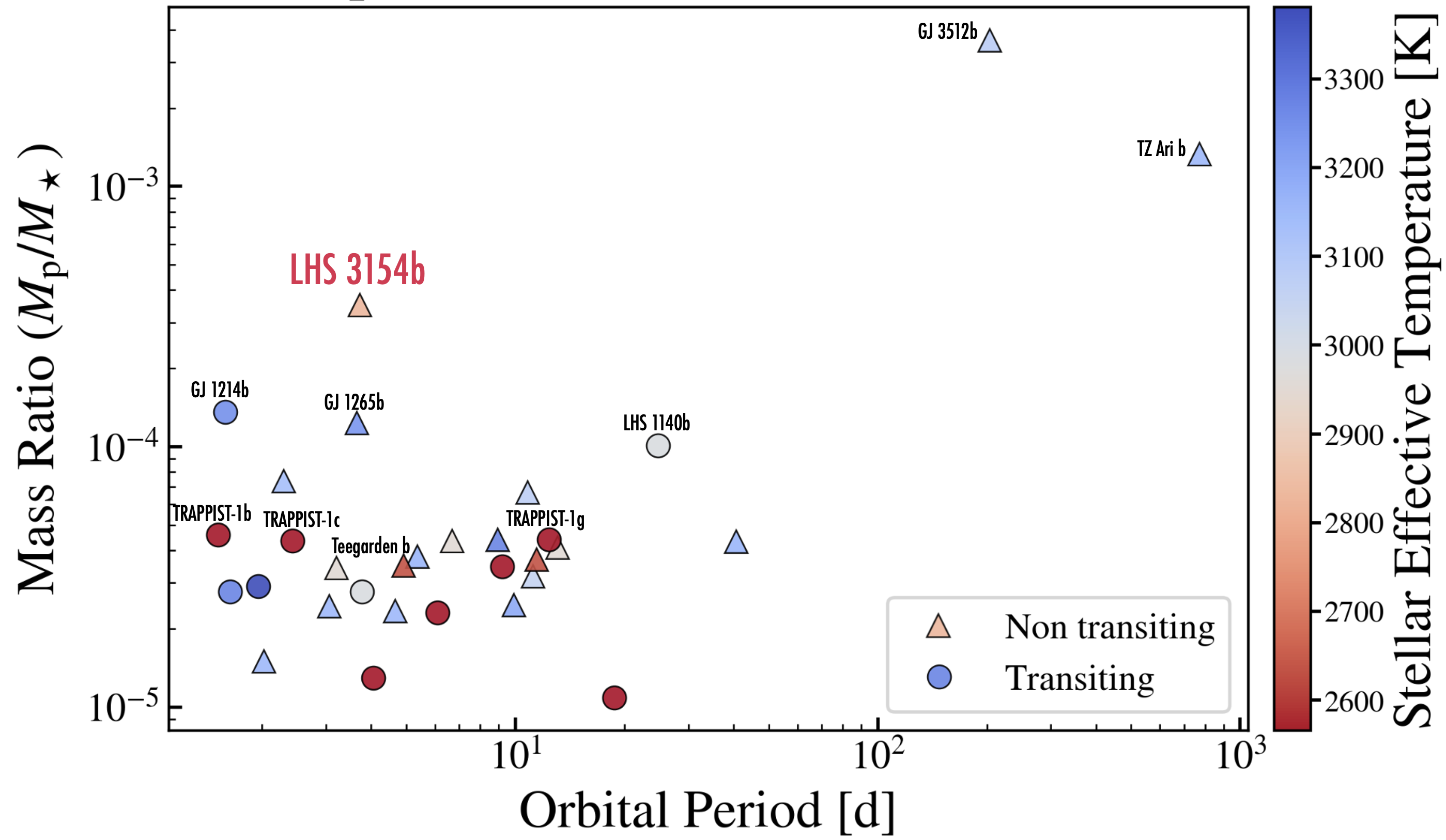}
\end{center}
\end{figure}
\noindent {\bf Fig~2.} Planet-to-star mass ratios for planets around ultracool stars ($M<0.25M_\odot$) for which precise mass measurements are available (with an uncertainty smaller than 30\%). Circles represent transiting planets, for which the planet's true mass is known. Triangles represent planets detected with the RV technique, for which only a lower mass limit ($M_p \sin i)$ is available. LHS 3154b, highlighted in red, is the most massive short-period planet orbiting an ultracool M dwarf.

\clearpage

\begin{figure}[t!]
\begin{center}
\includegraphics[width=\textwidth]{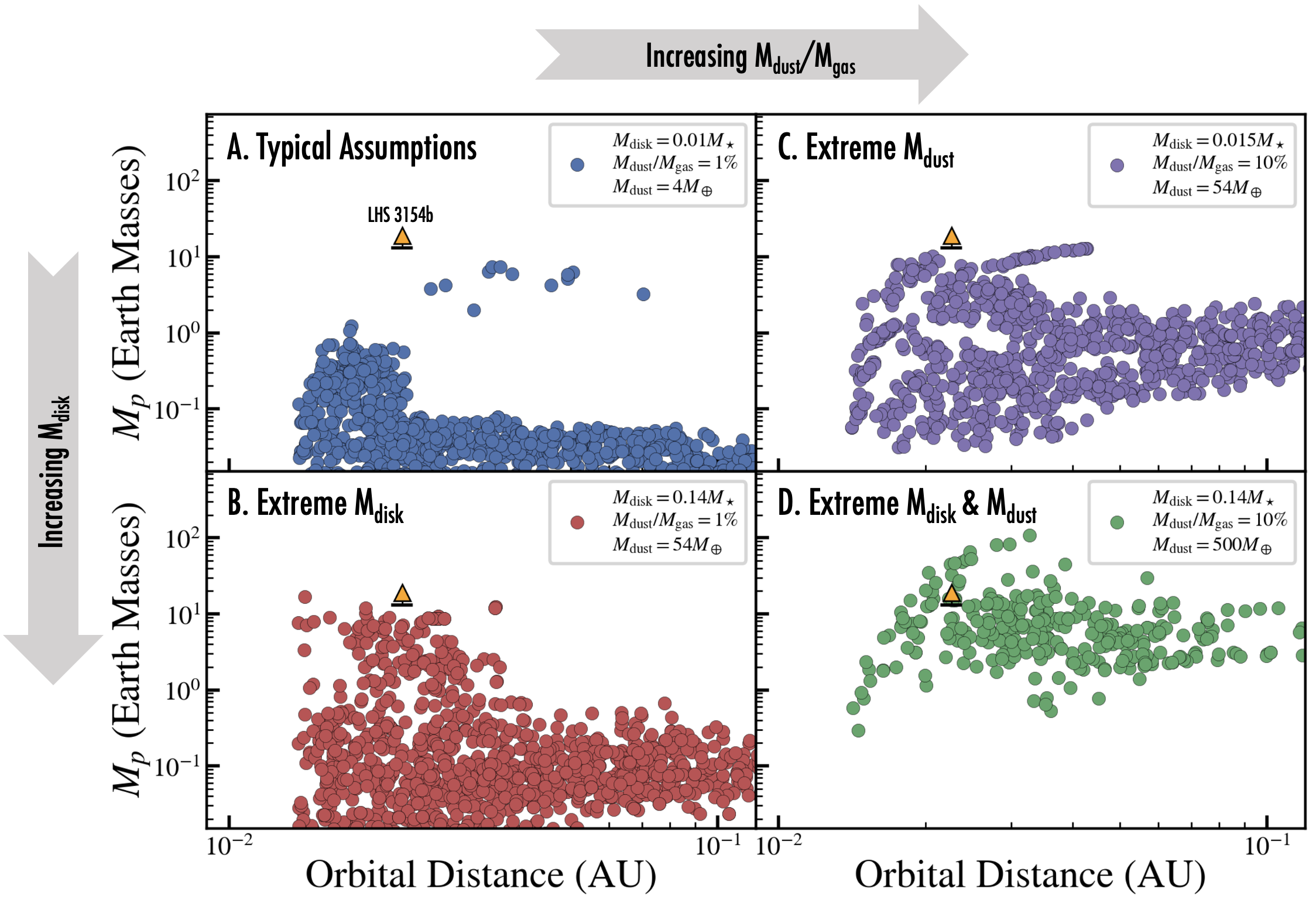}
\end{center}
\end{figure}
\noindent {\bf Fig~3.} Results from core-accretion planet formation models. Nominal current assumptions of planet formation (a) are incapable of forming planets as massive as LHS 3154b around $0.1M_\odot$ stars. LHS 3154b is denoted with yellow arrows where the base of the arrow denotes its minimum mass of 13.2 Earth masses. To form LHS 3154b-mass planets, we either need to increase the mass of the disk (b, d) or increase the dust to gas ratio (c, d), effectively increasing the amount of solid material available in the disk to form planetary cores.

\clearpage

\begin{table}[t!]
\centering
\begin{tabular}{cc}
\textbf{Stellar parameters}                                         & \textbf{Value}                                        \\ \hline
Right Ascension (J2016 epoch)                                       & 16:06:32.78                                           \\ 
Declination (J2016 epoch)                                           & +40:54:24.64                                          \\ 
Proper Motion (RA, $\mathrm{mas/yr}$)                               & $-716.302 \pm 0.036$                                  \\ 
Proper Motion (DEC, $\mathrm{mas/yr}$)                              & $162.486 \pm 0.046$                                   \\ 
Absolute Radial Velocity (km/s)                                     & $-42.05 \pm 0.34$                                     \\ 
Parallax ($\mathrm{mas}$)                                           & $63.423 \pm 0.034$                                    \\ 
Spectral type                                                       & M6.5                                                  \\ 
V-band magnitude (mag)                                              & $17.65 \pm 0.2$                                       \\ 
TESS magnitude (mag)                                                & $13.2266 \pm 0.007$                                   \\ 
J-band magnitude (mag)                                              & $11.05 \pm 0.018$                                     \\ 
$K_S$-band magnitude (mag)                                          & $10.072 \pm 0.019$                                    \\ 
Effective temperature (K)                                           & $2861 \pm 77$                                         \\ 
Metallicity ([Fe/H])                                                & Solar                                                 \\    
Mass ($M_\odot$)                                                    & $0.1118 \pm 0.0027$                                   \\ 
Radius ($R_\odot$)                                                  & $0.1405 \pm 0.0038$                                   \\ 
Luminosity ($L_\odot$)                                              & $0.00119_{-0.00014}^{+0.00015}$                       \\ 
Distance (pc)                                                       & $15.7531 \pm 0.0084$                                  \\ 
Space velocities (km/s)                                             & $U=-42.32\pm0.10$, $V=-54.53\pm0.21$, $W=4.16\pm0.25$ \\ 
Rotation period (days)                                              & $114\pm22$ (inferred)                                 \\ 
Age ($10^9$ years)                                                  & $5_{-2}^{+4}$                                         \\ 
H$\alpha$ emission ($\log L_{\mathrm{H\alpha}}/L_{\mathrm{bol}}$)   & $-5.3$                                                \\ 
Projected rotational velocity (km/s)                                & $<2$                                                  \\ \hline 
\textbf{Planet parameters}                                          & \textbf{Value}                                        \\
Orbital Period (days)                                               & $3.71778_{-0.00081}^{+0.00080}$                       \\  
Time of conjunction ($\mathrm{BJD_{TDB}}$)                          & $2458874.26_{-0.14}^{+0.14}$                          \\  
Time of periastron ($\mathrm{BJD_{TDB}}$)                           & $2458874.02_{-0.57}^{+0.67}$                          \\  
Eccentricity                                                        & $0.076_{-0.047}^{+0.057}$                             \\  
Argument of periastron (deg)                                        & $82_{-47}^{+102}$                                     \\  
Radial velocity semiamplitude (m/s)                                 & $23.4_{-1.4}^{+1.5}$                                  \\ 
Minimum mass $M \sin i$ ($M_\oplus$)                                & $13.15_{-0.82}^{+0.84}$                               \\ 
Orbital semimajor axis (au)                                         & $0.02262\pm0.00018$                                   \\\hline 
\end{tabular}
\end{table}
\noindent {\textbf{Table 1. Properties of LHS 3154 and its planet LHS 3154b.} Stellar parameters were derived using the \texttt{HPF-SpecMatch} code from HPF spectra, along with the relations of \cite{mann2015} and \cite{mann2019}. Planet parameters were derived from a nested sampling fit of the HPF RVs. Reported values are the median values along with the 68\% credible intervals.}


\baselineskip24pt

\section*{Supplementary materials}
Materials and Methods\\
Figs. S1 to S6\\
Tables S1 to S3\\
References \textit{(40-71)}

\section*{Materials and Methods}

\noindent\underline{Spectroscopic observations and analysis}\\
We obtained high resolution spectra of LHS 3154 with the Habitable-zone Planet Finder (HPF) spectrograph \cite{mahadevan2012,mahadevan2014}. HPF is a milli-Kelvin temperature stabilized \cite{stefansson2016} fiber-fed \cite{kanodia2018fiber} near-infrared spectrograph on the 10m Hobby-Eberly Telescope \cite{ramsey1998,hill2021} in Texas covering the $z$, $Y$ and $J$ bands from 810-1280nm at a resolving power $R=55,000$. All observations were conducted as part of the HET queue \cite{shetrone2007}. In total, we obtained 137 spectra in 69 visits. In each visit, we obtained two spectra with an exposure time of 969s. Across all visits, the median S/N per spectrum is 52.7 per 1D extracted pixel at $1\mathrm{\mu m}$. We removed 5 spectra with a low S/N < 15. After performing this cut, we have 132 spectra in 67 visits with a median S/N of 53.4 per 1D extracted pixel at $1\mathrm{\mu m}$. After binning the spectra per HET visit ($\sim$30 min bins), we obtain a median RV uncertainty of 7.3 m/s. We used the binned data for subsequent analysis, which has a time baseline of 810 days from January 23 2020 to April 13 2022. Tables S1 and S2 list the RVs from HPF and associated activity indicators derived from the HPF spectra used in this work, respectively.

HPF has a NIR Laser Frequency Comb (LFC) calibrator to provide a precise wavelength solution and track instrumental drifts, which has been shown to enable $\sim$20 $\mathrm{cm/s}$ RV calibration precision in 10 minute bins \cite{metcalf2019}. Following \cite{stefansson2020}, we elected not to use the simultaneous LFC calibration during the observations to minimize the risk of contaminating the science spectrum from scattered light from the LFC, and the drift correction was performed as described in \cite{stefansson2020}, which has been shown to enable precise wavelength calibration at the $\sim$$30~\mathrm{cm/s}$ level.

The HPF 1D spectra were reduced using the HPF pipeline, following the procedures in \cite{ninan2018}, \cite{kaplan2018}, and \cite{metcalf2019}. Following the 1D spectral extraction, we reduced the HPF radial velocities using an adapted version of the \texttt{SERVAL} (SpEctrum Radial Velocity Analyzer) pipeline \cite{zechmeister2018}, which is described in \cite{stefansson2020}. In short, \texttt{SERVAL} uses the template matching algorithm to derive RVs, which has been shown to be particularly effective at producing precise radial velocities for M-dwarfs. \texttt{SERVAL} uses the \texttt{barycorrpy} package \cite{kanodia2018}, which uses the methodology of \cite{wright2014} to calculate accurate barycentric velocities. Following \cite{stefansson2020}, we only use the 8 HPF orders that are cleanest of tellurics, covering the wavelength regions from 8540-8890\AA, and 9940-10760\AA. We subtracted the estimated sky background from the stellar spectrum using the dedicated HPF sky fiber. We explicitly masked out telluric lines and sky-emission lines to minimize their impact on the RV determination. 

We modeled the RVs using \texttt{juliet} \cite{espinoza2019}, which uses the \texttt{radvel} code \cite{fulton2018} to calculate the Keplerian model. We used the \texttt{dynesty} nested sampler \cite{speagle2019} available in \texttt{juliet} to perform a dynamic nested sampling of the posteriors. We modeled the Keplerian using the usual prescription of orbital period ($P$), time of inferior conjunction ($T_{\mathrm{conj}}$), eccentricity ($e$), argument of peristron ($\omega$), and RV semi-amplitude ($K$). We account for any additional noise not accounted for the errors with a white noise jitter term. To account for any potential long-period outer companions, we additionally experimented with adding a slope to the RV model. The resulting posterior on the slope parameter was fully consistent with no slope over the observing baseline, and we therefore adopt a model without a slope. Table S3 summarizes the priors and the posteriors from the RV fit. 

\vspace{0.8cm}
\noindent\underline{Spectroscopic stellar activity analysis}\\
To study the activity of the star, we measured a number of stellar activity indicators from the HPF spectra following the procedures in \cite{zechmeister2018} and \cite{stefansson2020b}. Figure S1 shows the Generalized Lomb-Scargle (LS) periodograms of the RVs along with different activity indicators measured from the HPF spectra, including: the RV residuals, the Differential Line Width (dLW), the Chromatic Index (CRX), and line indices of the three Calcium II Infrared Triplet (Ca II IRT) lines. To calculate the LS periodograms, we used the periodogram functions in the \texttt{astropy.timeseries} package, and the false alarm probabilities\footnote{Although the False Alarm Probability is commonly used in periodogram analysis in radial velocity data, it has known limitations (see e.g., discussion in \cite{fischer2016}.)} were calculated using the \texttt{bootstrap} method implemented in this same package. Additionally, in Figure S1, we show the Window Function (WF) of our RV observations. All of the periodograms in Figure S1 are normalized assuming Gaussian noise, except the window function is normalized such that the highest peak has a power of 1. We see no peaks in the activity indicators or the window function at 3.7 days. Coupled with the strong RV signal, this strongly suggests that the RV variations are not due to stellar activity.

Additionally, we obtained an $R=2,500$ optical spectrum obtained with LRS2 \cite{chonis2016} on the Hobby-Eberly Telescope to measure the pseudo-equivalent width of the H$\alpha$ line, pEW(H$\alpha$), which we then use to estimate the fractional H$\alpha$ luminosity to the overall bolometric luminosity $\log (L_{H\alpha}/L_{bol})$. We obtained a spectrum with the LRS2-R unit in the `red' channel from 6450-8400\AA. Prior to measuring pEW(H$\alpha$), we shift the spectra to zero radial velocity, accounting for the barycentric velocity, and absolute velocity of the star. The LRS2 spectrum was reduced using the Panacea pipeline \cite{zeimann}, which performs basic CCD reduction tasks, wavelength calibration, fiber extraction, sky subtraction and flux calibration. We measure the pEW(H$\alpha$) using the following equation:
\begin{equation}
\mathrm{pEW(H\alpha)} = \int_{\lambda_1}^{\lambda_2} \left( 1 - \frac{F(\lambda)}{F_{pc}}\right) d\lambda
\end{equation}
where we integrate over the limit from $\lambda_1 = 6560$\AA\ and $\lambda_2 = 6566$\AA. $F_{pc}$ is the average of the median flux in the pseudo-continuum in the ranges from $6545-6559$\AA, and $6567-6580$\AA\ after removing a linear slope fit to that range seen in the pseudo-continuum surrounding the H$\alpha$ line for late M dwarfs.

Figure S2 shows the spectrum of LHS 3154 surrounding the H$\alpha$ line compared to an LRS2 spectrum of the active M8 star VB 10 \cite{kanodia2022}. VB 10 shows clear H$\alpha$ emission, while we see no H$\alpha$ emission in the LHS 3154 spectrum. For LHS 3154, we measure $\mathrm{pEW(H\alpha)}=-0.31$\AA. Accounting for the S/N of the observation, we obtain a statistical uncertainty of $0.1$\AA.

To estimate $\log (L_{H\alpha}/L_{bol})$, we use the following equation,
\begin{equation}
\log \left( \frac{L_{H\alpha}}{L_{bol}}\right) = \log \chi + \log(-\mathrm{pEW(H\alpha})),
\end{equation}
where $\log \chi$ is the ratio of the flux in the continuum near H$\alpha$ to the bolometric flux. We use the $\chi(T_{\mathrm{eff}})$ function from \cite{reiners2008}, who estimated the functional form of $\chi$ as a function of $T_{\mathrm{eff}}$ using PHOENIX spectra. We obtain a $\log(\chi)=-4.8$, suggesting a $\log (L_{H\alpha}/L_{\mathrm{bol}})=-5.3$. This agrees with the value reported by \cite{metodieva2015} of $\log (L_{\mathrm{H\alpha}}/L_{\mathrm{bol}}) \leq -5.21$ for LHS 3154. Judging from the sample presented in \cite{west2015}, these values of $\log (L_{\mathrm{H\alpha}}/L_{\mathrm{bol}})$ are suggestive of an exceptionally quiet ultracool star whose rotation period should be in excess of 100 days.

\vspace{0.8cm}
\noindent\underline{Photometric observations and analysis}\\
LHS 3154 was observed by the Transiting Exoplanet Survey Satellite (TESS) \cite{ricker2015} in 3 sectors in Sector 24 (April 16, 2020 -- May 13, 2020), Sector 25 (May 13, 2020 -- June 8, 2020), and Sector 51 (April 22, 2022 -- May 18, 2022). The TESS photometry of LHS 3154 has minimal dilution from nearby stars with a low contamination ratio of 0.0082.

To check if TESS reveals periodicities close to the period of LHS 3154b, we used the Systematics-Insensitive Periodogram (SIP) package \cite{hedges2020}. SIP is capable of creating periodograms while simultaneously accounting for TESS systematics within and across TESS sectors. To do so, SIP uses the TESS Target Pixel File data and apertures assigned by the TESS pipeline to build simple aperture photometry light curves of a given target while building an estimate of the background contribution. To build a periodogram, SIP simultaneously fits systematic regressors and sinusoidal components to the time-series data, including regularization terms to avoid overfitting. Figure S3 shows the resulting corrected TESS photometry along with corresponding SIP periodograms for all sectors together, along with periodograms for individual sectors. We see no periodicity at the period of the planet. 

To create the SIP periodogram, we elected to use all data irrespective of quality flag, as during testing, we observed this resulted in smoother SIP-corrected light curves. No peaks were seen at the planet period when removing data that had non-zero TESS quality flags. Additionally, we also tried generating Generalized-Lomb Scargle periodograms of the PDCSAP light curves for each sector, which revealed no consistent peaks at the planet period. 

Given the close-in orbit of the planet, the planet has a relatively high geometric transit probability of $R_\star/a = 2.9\%$. Using the ephemerides from the HPF RVs, we looked for evidence of transits in the TESS data. Figure S3d shows the expected transit time using the RV ephemerides. The model in Figure S3d assumes a most likely radius of the planet of $R=3.65 R_\oplus$ using the \texttt{Forecaster} package \cite{chen2017}. We see that the TESS data confidently rules out any transits. 

Additionally, the Zwicky Transient Facility (ZTF) observed LHS 3154 in three different filters, $z_g$, $z_r$, and $z_i$, covering a baseline of 796, 788, and 814 days, respectively. All filters had a median exposure time of 30s. We removed 17, 6 and 4 points as high ($>1.9$) airmass points for the three filters, respectively. We additionally removed 12, 20, and 6 points as $>4\sigma$ outliers, leaving a total of 681, 695 and 152 points for periodogram analysis for the three filters, respectively. Figure S4 shows the ZTF photometry along with the associated generalized Lomb-Scargle periodograms for each individual filter. We see no peaks at the planet period in any of the three filters. In $z_g$, we see peaks at 28.4 days and half of that (14.3 days), which we attribute to the impact of the moon on the faint star in the bluest filter, where the impact of scattered light from the moon is expected to be the highest. In $z_r$ we see a peak at 38.7 days above the 0.1\% false alarm level, which we attribute to be a third of the $P=116$-day peak seen in the $z_i$ filter. The 116-day peak seen in the $z_i$ filter, and the two broad peaks seen in the $z_g$ filter at $\sim$90-140 days, are fully consistent with the expected rotation period value of $P=114\pm22$days as calculated by the relations from \cite{newton2017} for inactive ultracool M dwarfs. Given the broad array of peaks seen in the $z_g$, and $z_i$ filters from $\sim$90-140 days, we adopt a rotation period of $114\pm22$ days to encompass these possible peaks as being the true rotation period.

\vspace{0.8cm}
\noindent\underline{Starspot False Positive Analysis}\\
While our analyses of photometry and $L_{\mathrm{H}\alpha}$/$L_{\mathrm{bol}}$ of LHS 3154 indicate the stellar rotation period should be significantly longer than the orbital period of its exoplanet candidate, we nonetheless built a simple model to evaluate the possibility that the observed RV signal could be caused by starspots rotating across the stellar surface. We created our starspot model using the SOAP 2.0 \cite{dumusque2014} software package, which simulates the photometric and RV signals created by starspots and plage of various configurations.

Our SOAP 2.0 models adopt the same stellar parameters listed in Table 1, except that the rotation period is assumed to be 3.7 days---matching the period of the RV signal---and the stellar inclination angle $i$ is set to $90^{\circ}$.  We generated starspot signals for a single equatorial spot of varying size and spot-photosphere temperature difference ($\Delta T$).

Across a range of starspot size/temperature combinations, we found no model that reproduced the high-amplitude RV signal alongside a photometric modulation amplitude consistent with the TESS lightcurve.  For a $\Delta T = 600$ K---considered to be within the typical range for M dwarf spots \cite{reiners2010}---we find that a spot radius $R = 0.15R_*$ is required to match the observed RV amplitude. However, assuming non-polar stellar inclinations such a spot should create photometric modulation with amplitude of $\sim$2\%, which is easily ruled out by the TESS lightcurve. In combination with the collection of evidence suggesting LHS 3154 is especially old and slowly rotating, these results indicate that conventional starspot modulation cannot easily explain the combination of RV and photometric signals observed for this star.

\vspace{0.8cm}
\noindent\underline{Age}\\
Although ages of M dwarfs are difficult to constrain precisely, to estimate a likely age for LHS 3154 we follow \cite{newton2016}, which show slowly rotating inactive M dwarfs with $0.1 < M < 0.25 M_\odot$ have an average age of $5_{-2}^{+4}$ GYr. We adopt this value as our most likely age for LHS 3154. This old age agrees with the Galactic space velocities of LHS 3154, from which we follow \cite{carrillo2020} to calculate membership probabilities of 95\%, 4.8\%, and $\ll$1\% for LHS 3154 to be a member of the Galactic thin-disk, thick-disk, and Galactic halo populations, respectively. From the Galactic velocities, we used the BANYAN $\Sigma$ tool \cite{gagne2018}, which rules out a membership to 27 well-characterized young associations within 150 pc with 99.9\% confidence.

\vspace{0.8cm}
\noindent \underline{Constraints on additional planets}\\
We looked for evidence of additional planets in the system. A Box-Least-Squares search of the TESS data reveals no transiting planets. Additionally, a periodogram analysis of the HPF RV residuals (see Figure S1b) reveals no additional significant peaks. To place constraints on the upper mass limits of possible planets in the system, we performed an additional fit of the HPF RVs assuming a 2 planet model in \texttt{juliet}, assuming both planets are on circular orbits. We placed broad priors on the second planet, assuming a period from 4 to 50 days. Figure S5 shows the results in the $m_c\sin i$ vs orbital period plane. This places an upper limit of the mass of a hypothetical planet c of $m_c \sin i < 8.3M_\oplus$ at $3\sigma$ confidence for periods between $4-50$ days. A similar fit investigating the possibility of an inner planet from 0.5 to 3.6 days also revealed no clear signatures of another planet.

We additionally investigated the possibility of planets in 2:1 (1.86 days) or 1:2 (7.43days) inner/our mean motion resonances (MMR) similar to the case seen in the M dwarf system GJ 876 \cite{marcy2001}. We see no significant evidence for RV signals at those periods, with joint fits of the known 3.7day signal and informative priors constrained at the 2:1 or 1:2 MMR periods returning RV semi-amplitudes that are consistent with a flat line at $2\sigma$. All signals were assumed to be circular. We place formal $3\sigma$ upper mass limits of $<2.5M_\oplus$ and $<5.0M_\oplus$ for the 1.86day and 7.43day MMR case, respectively. Given the difficulty of recognizing planetary systems in a 2:1 MMR from the present dataset, we recommend long-term RV monitoring of LHS 3154 to search for evidence of any possible orbital precession.

\vspace{0.8cm}
\noindent\underline{Planet formation models}\\
We use the semi-analytical planet formation model of \cite{miguel2020}, based on the core accretion scenario and assuming that planets grow initially through the accretion of planetesimals. We present here some details on our model for completeness, but we refer the reader to \cite{miguel2020} for more details and equations used in the calculations. 

For each planetary system, we start the calculations assuming a disk of gas and solids and a number of initial embryos growing in the midplane. Each initial embryo has a radius of 500 km, corresponding to a mass of approximately $1.4\times10^{-4}$ Earth masses and the total number of initial embryos in each system is a few hundred with the exact number depending on the mass of the host star and disk. For example, for the nominal case of M$_{disk}=0.01$M$_{\star}$, we have approximately 600 initial embryos in the disk, which adds 0.08 Earth masses (or $2.38\times10^{-7}$  M$_{\odot}$) of solids to the system. We note that this is two orders of magnitude lower than the total mass of solids in the disk. 

The embryos grow with a timescale that is determined by the velocity dispersion of the swarm of planetesimals in its surroundings. In order to model the formation of a Neptune-like planet, we included gas accretion in our calculations. Following \cite{ida2013}, we start the gas accretion once the planet has a mass larger than the critical mass,
\begin{displaymath}
M_{crit} = 10 \Big( \frac{\dot{M_s}}{10^6 \ M_{\oplus} \  yr^{-1}}\Big)^{0.25} M_{\oplus},
\end{displaymath} 
where $\dot{M_s}$ is the accretion rate of solids. The gaseous envelope is accreted on a Kelvin–Helmholtz time-scale,
\begin{displaymath}
\frac{dM_p}{dt}=\frac{M_p}{\tau_{KH}},
\end{displaymath}
where M$_p$ is the planet’s mass and $\tau_{KH}$ is the Kelvin–Helmholtz time-scale given by,
\begin{displaymath}
\tau_{KH} = 10^9 \Big(\frac{M_p}{M_{\oplus}}\Big)^{-3}\ \mathrm{yr}.
\end{displaymath}
Planets stop the gas accretion process when gas dissipates from the disk or when the planet’s Hill radius becomes larger than two times the disk scale height, a result based on detailed numerical calculations (see \cite{ida2013} and references therein). The planets do not remain in their initial orbits, and migrate with type I migration due to an imbalance in the corotation and Lindblad torques, and this switches to type II migration once the mass of the planet is big enough to allow the opening of a gap in the orbit. In our calculations, we slow down migration randomly by multiplying the rate of change of the semimajor axis by a constant value between 1 and 0.1, to take into account the uncertainties in these processes \cite{ida2004,miguel2020}. Planets can additionally be trapped in resonances, and may undergo orbit crossings and possible mergers after the gas has dissipated from the disk. Each planet formation model is run for $10^8$ years.

The solid dust disks and gaseous disks evolve with time in our calculations. The solid disk evolves due to the accretion of the planetary embryos, which slowly decreases the amount of solids in the planets feeding zones. More importantly, the dust is also depleted globally due to the gas drag effect. Planetesimals orbit at a Keplerian speed and suffer from a drag force caused by the gas moving at sub-Keplerian speed. This causes the planetesimals to drift towards the star and deplete the disk of solids quite efficiently \cite{mosqueira2003, miguel2016}. The gas disk also evolves locally and globally: locally due to the gas accretion onto the embryos, and globally it decreases exponentially with a dissipation time-scale which we vary between $10^6$ and $10^7$ years. 

The initial conditions of the protoplanetary disk strongly influence the outcomes of the planet formation simulations, in particular the final masses and architectures of the planetary systems \cite{miguel2011a, mordasini2012}. Observations of protoplanetary disks suggest a relation between the mass of the star and that of the protoplanetary disk of $M_{\mathrm{disk}}/M_{\star}=0.01$ \cite{williams2011}. Subsequent studies showed that this relation also depends on the age of the disk \cite{pascucci2016,ansdell2016}, but for young disks of a few million years, the relation M$_{disk}$/M$_{\star}=0.01$ is still valid considering the large dispersion in the observational data. Moreover, a recent study \cite{manara2022} showed that for stars of 0.1 M$_{\odot}$, most of the observed disks have masses between 0.01 and 0.001 times that of the star. We use a relation of M$_{disk}$/M$_{\star}=0.01$ as the nominal case, and explore extreme cases based on protoplanetary disk observations in order to explain the formation of LHS 3154b.

In total, we run four different simulation scenarios, all of which consist of 300 planetary systems that are considered stable against gravitational collapse by the Toomre Q stability criterion. The initial conditions of the four disks considered are summarized in Figure S6, and the outcomes of the planet formation simulations are summarized in Figure 3. For the first scenario, we adopt the nominal relation between the mass of the star and the mass of the disk, and we assume a $M_{\mathrm{dust}}/M_{\mathrm{gas}}=1\%$, typical of the value seen in the ISM \cite{bohlin1978}. This model is shown in Figure 3a. This model is capable of creating planets larger than Mars and up to a few Earth-mass planets, similar to the TRAPPIST-1 planets. However, it is not capable of creating planets as massive as LHS 3154b.

To create more massive planets, we explored increasing the mass of the protoplanetary disk. In Figure 3b, we increase the mass of the protoplanetary disk to the maximum dust-mass value consistent with observations of protoplanetary disks from \cite{pascucci2016} for the lowest mass stars of $0.1M_\odot$ with $M_{\mathrm{dust}} = 54M_\oplus$. Assuming a $M_{\mathrm{dust}}/M_{\mathrm{gas}}=1\%$, this corresponds to a disk mass of $M_{\mathrm{disk}} = 0.14M_\star$. From Figure 3b, we see that this model is barely capable of creating planets that have similar masses and orbital distances as LHS 3154b. 

In addition to exploring changing the protoplanetary disk mass, we explored changing the gas-to-dust ratio. A $M_{\mathrm{dust}}/M_{\mathrm{gas}}=1\%$ is typically assumed to derive total gas disk masses from measured dust disk masses. However, the mass of dust derived from protoplanetary disk observations is only sensitive to grain sizes up to $\sim$cm, and is based on the emission from regions with r$>$10 AU \cite{bergin2018, manara2018}. Therefore, we take dust masses derived from observations as a minimum value in our calculations and explore different $M_{\mathrm{dust}}/M_{\mathrm{gas}}$ relations. In Figure 3c, we also show the limiting case at the upper limit of dust measurements in protoplanetary disks, but now assuming a $M_{\mathrm{dust}}/M_{\mathrm{gas}}=10\%$. This corresponds to a protoplanetary disk mass of $M_{\mathrm{disk}}=0.015M_\star$. From Figure 3b, we see that this model is also capable of creating planets that have similar masses and orbital distances as LHS 3154b.

We additionally note that for both the models in Figure 3b and 3c, when a massive planet is formed, it is the only massive planet formed in the system. During their formation, these planets accrete solids more efficiently and become an oligarch among the group of planet embryos, growing faster than the others and merge with the remaining embryos. This result obtained in the simulations also agrees with what we see in the LHS 3154 system: we see no evidence of other massive planets in the system.

Lastly, in Figure 3d, we keep the mass of the disk the same as in Figure 3b, while assuming the same $M_{\mathrm{dust}}/M_{\mathrm{gas}}=10\%$ as in Figure 3c. As expected, this model has the highest disk mass and the most amount of dust available ($M_{\mathrm{dust}}=500M_\oplus$), and is capable of creating the most massive planets. However, we note that this model would imply dust masses that are a factor of 10 larger than that those seen by protoplanetary disk observations \cite{pascucci2016,ansdell2016}. As such, we prefer the models in Figure 3b and 3c over the model in 3d, even if they marginally form planets such as LHS 3154b.

Disks are complex systems, and the disks around low mass stars are difficult to observe and are poorly understood. Along with detecting new planets around the lowest mass stars, further observations of protoplanetary disks at the lowest mass end will help improve our understanding of planet formation around the lowest mass stars.

\clearpage

\begin{figure}
\begin{center}
\includegraphics[width=0.75\columnwidth]{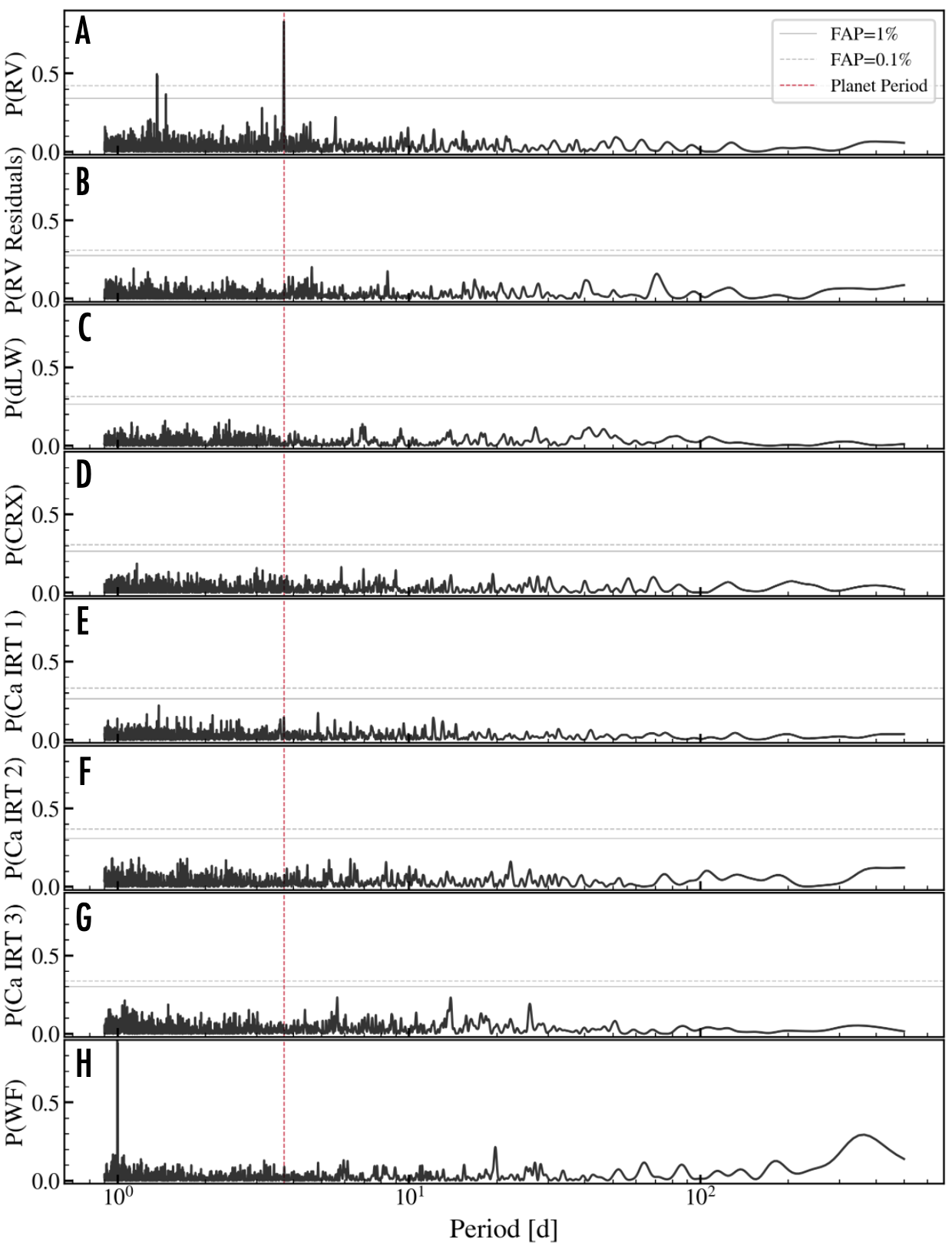}
\end{center}
\end{figure}
\noindent {\bf Fig~S1.} Lomb-Scargle periodograms of the HPF RVs along with different activity indicators. The planet period ($P=3.7$days) is highlighted with the dashed red line. False alarm probabilities of 1\% and 0.1\% calculated using a bootstrap method are denoted with the grey solid and grey dashed lines, respectively. \textbf{(A)} Periodogram of the HPF RVs. \textbf{(B)} Periodogram of the HPF RV residuals after subtracting the best-fit RV model. \textbf{(C)} Periodogram of the Differential Line Width (dLW) indicator. \textbf{(D)} Periodogram of the Chromatic Index activity indicator. \textbf{(E-G)} Periodograms of the Ca II IRT indices for the three Ca II IRT lines. \textbf{(H)} The window function of the HPF RVs, showing a clear sampling peak at 1 day. The power in \textbf{(A-G)} is normalized using the formalism in \cite{zechmeister2009}, and \textbf{(H)} is normalized so that the highest peak is unity.

\clearpage

\begin{figure}[t!]
\begin{center}
\includegraphics[width=\columnwidth]{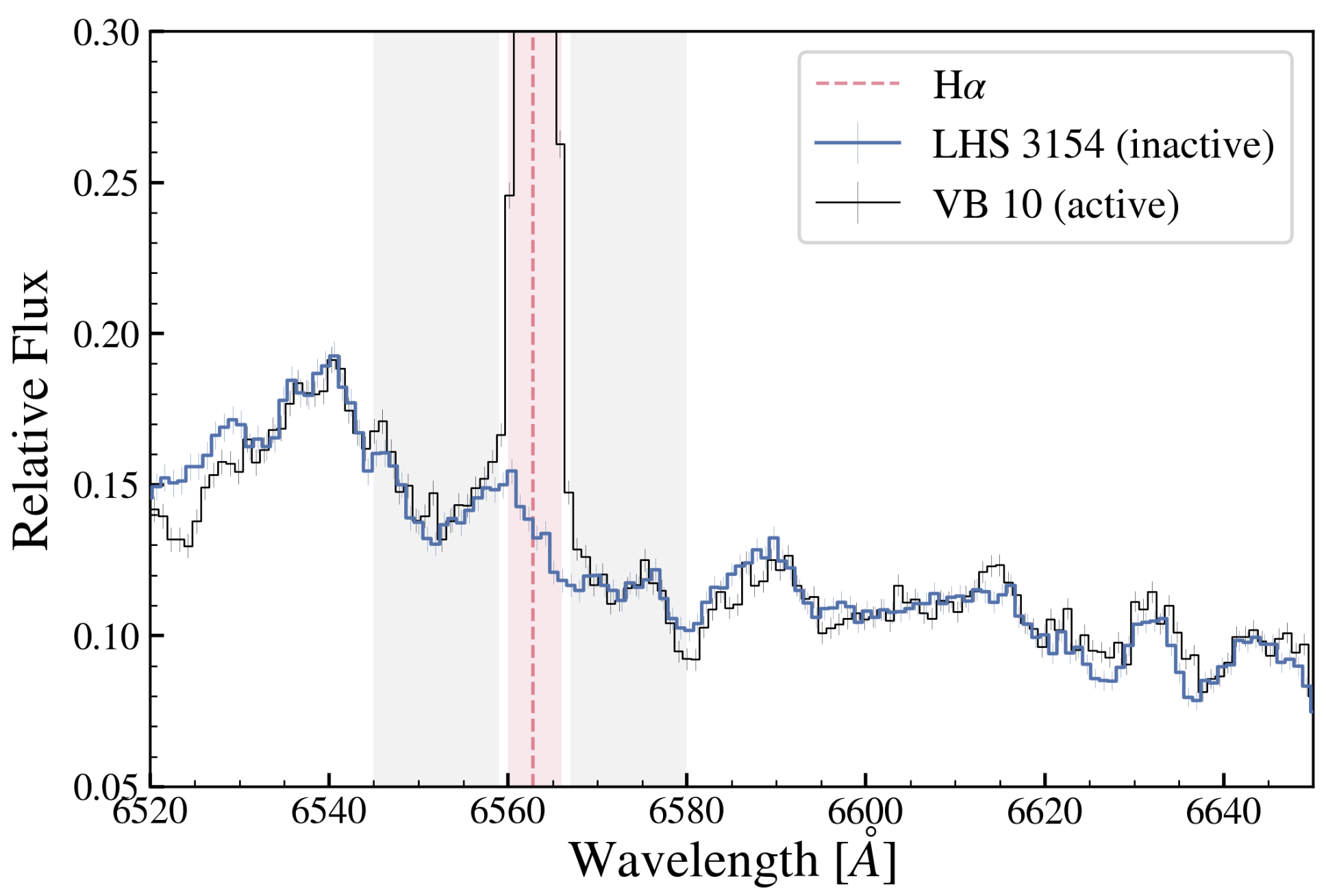}
\vspace{-0.5cm}
\end{center}
\end{figure}
\noindent \textbf{Fig~S2.} LRS2 spectroscopic observations of the H$\alpha$ line region of LHS 3154 (blue) and of the active M8 star VB 10 (black), which shows clear H$\alpha$ emission, for comparison. We do not see any signs of H$\alpha$ emission in LHS 3154, indicating it is an inactive star. The red vertical line denotes the location of the H$\alpha$ line, and the grey areas denote the pseudo-continuum comparison regions we used.

\clearpage

\begin{figure}[t!]
\begin{center}
\includegraphics[width=\textwidth]{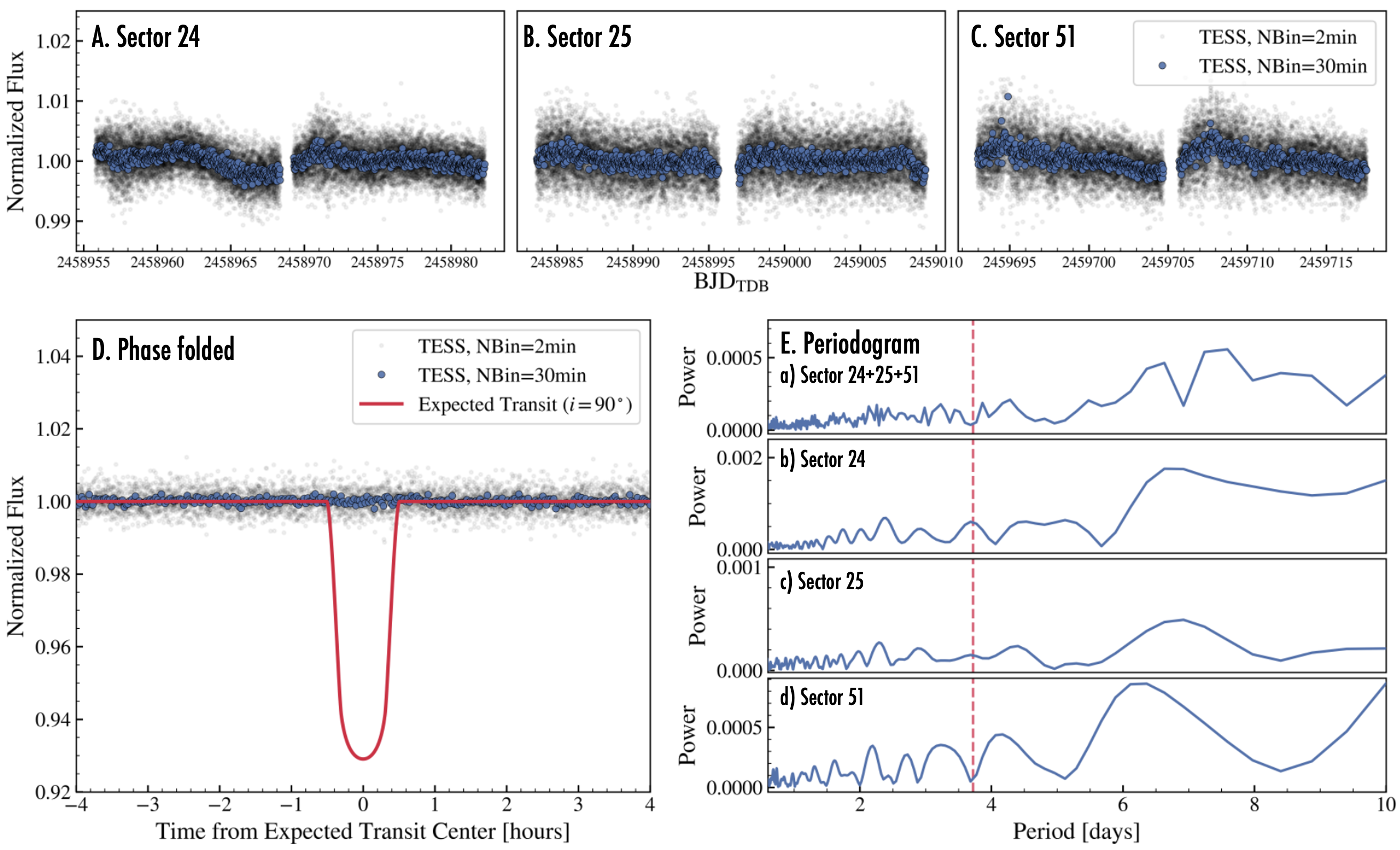}
\end{center}
\end{figure}
\noindent \textbf{Fig~S3.} TESS Photometry in 2min (black points) and 30min bins (blue points) from \textbf{(A)} Sector 24, \textbf{(B)} Sector 25, and \textbf{(C)} Sector 51. \textbf{(D)} TESS photometry phase folded at the expected transit from the RV ephemerides, along with an expected transit model in red assuming a transiting edge-on orbit ($i=90^\circ$) and a nominal expected radius of $3.65 R_\oplus$. The TESS data confidently rule out any transits. \textbf{(E)} Systematics Insensitive Periodograms of \textbf{(a)} Sector 24, 25, and 51 together, \textbf{(b)} Sector 24, \textbf{(c)} Sector 25, and \textbf{(d)} Sector 51, individually. No periodic variations are seen at the planet period (red vertical dashed line).

\clearpage

\begin{figure}[t!]
\begin{center}
\includegraphics[width=\textwidth]{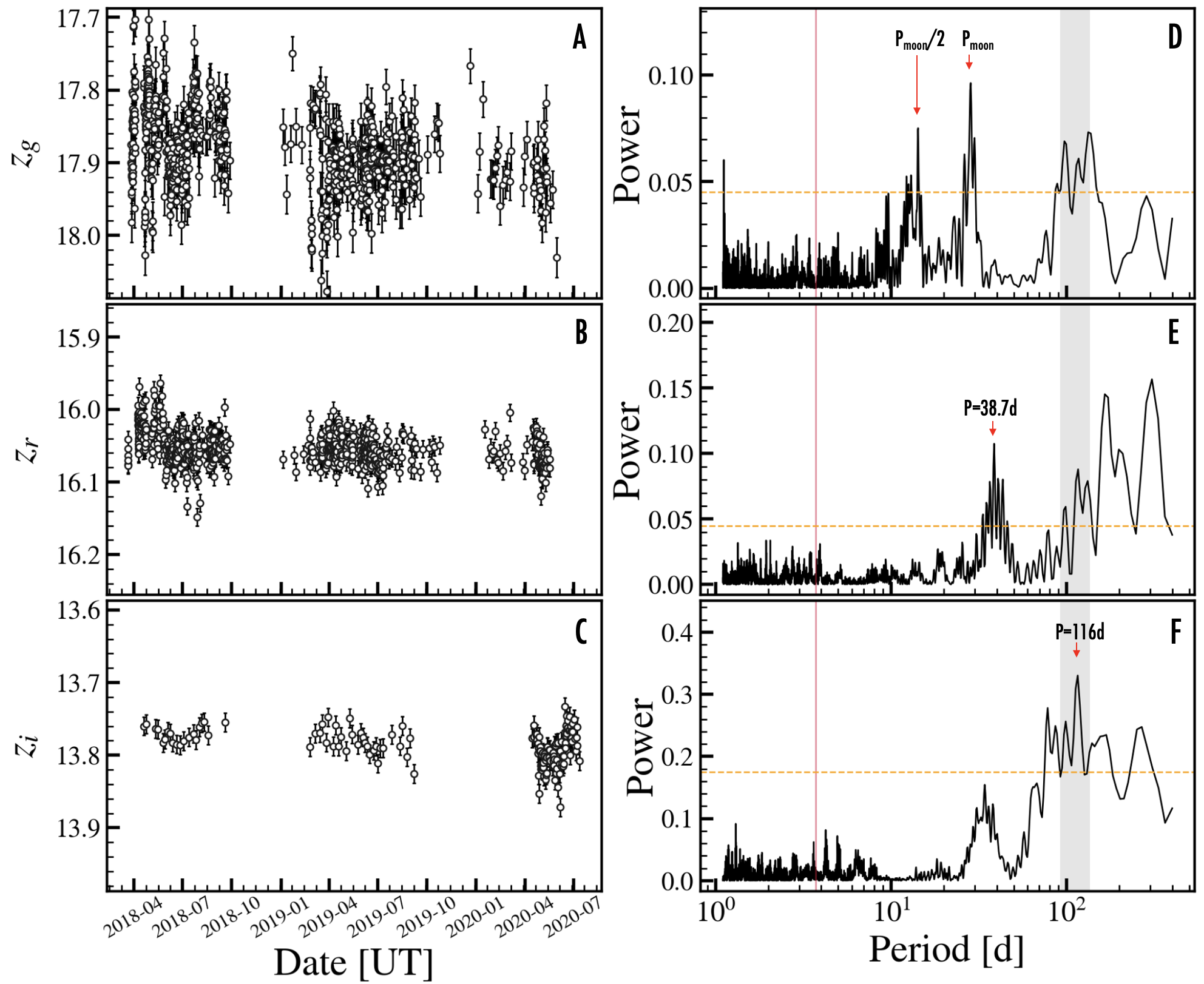}
\end{center}
\end{figure}
\noindent \textbf{Fig~S4.} Photometry from ZTF in three filters: \textbf{(A)} $z_g$, \textbf{(B)} $z_r$, and \textbf{(C)} $z_i$, along with associated generalized Lomb-Scargle periodograms \textbf{(D-F)}. The red vertical line denotes the 3.7 day planet period. The horizontal orange line denotes the $0.1\%$ false alarm probability threshold. We see no significant peaks at the planet period in any filters. The expected $1\sigma$ range of the stellar rotation period of $P=114\pm22$ days calculated from the relations in \cite{newton2017} is shown with the grey shaded regions. In $z_i$, we see a peak within that region at $P=116$ days. We also see a peak at 38.7days in $z_r$, which we attribute as being a third of the $P=116$-day peak seen in the $z_i$ filter. We adopt a rotation period of $P=114\pm22$ days to encompass these peaks.

\clearpage

\begin{figure}[t!]
\begin{center}
\includegraphics[width=\columnwidth]{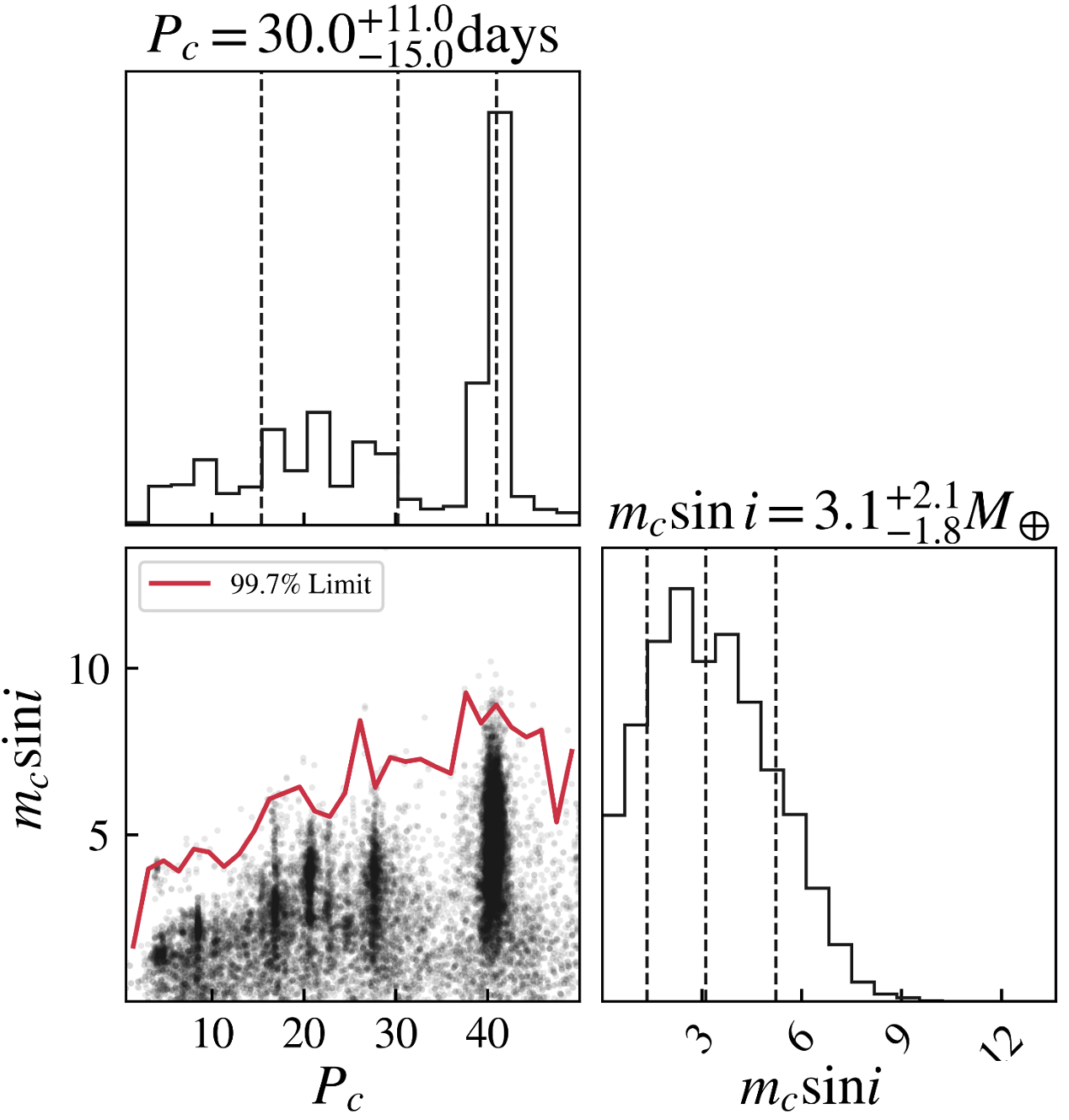}
\vspace{-0.5cm}
\end{center}
\label{fig:rv2}
\end{figure}
\noindent \textbf{Fig~S5.} Constraint of a second planet in the system. We do not see clear evidence of another planet on short period orbits from 4-50 days. Additional test runs for inner planets from 0.5 - 3.6days also reveal no clear signatures of an inner planet.

\clearpage

\begin{figure}[t!]
\begin{center}
\includegraphics[width=\columnwidth]{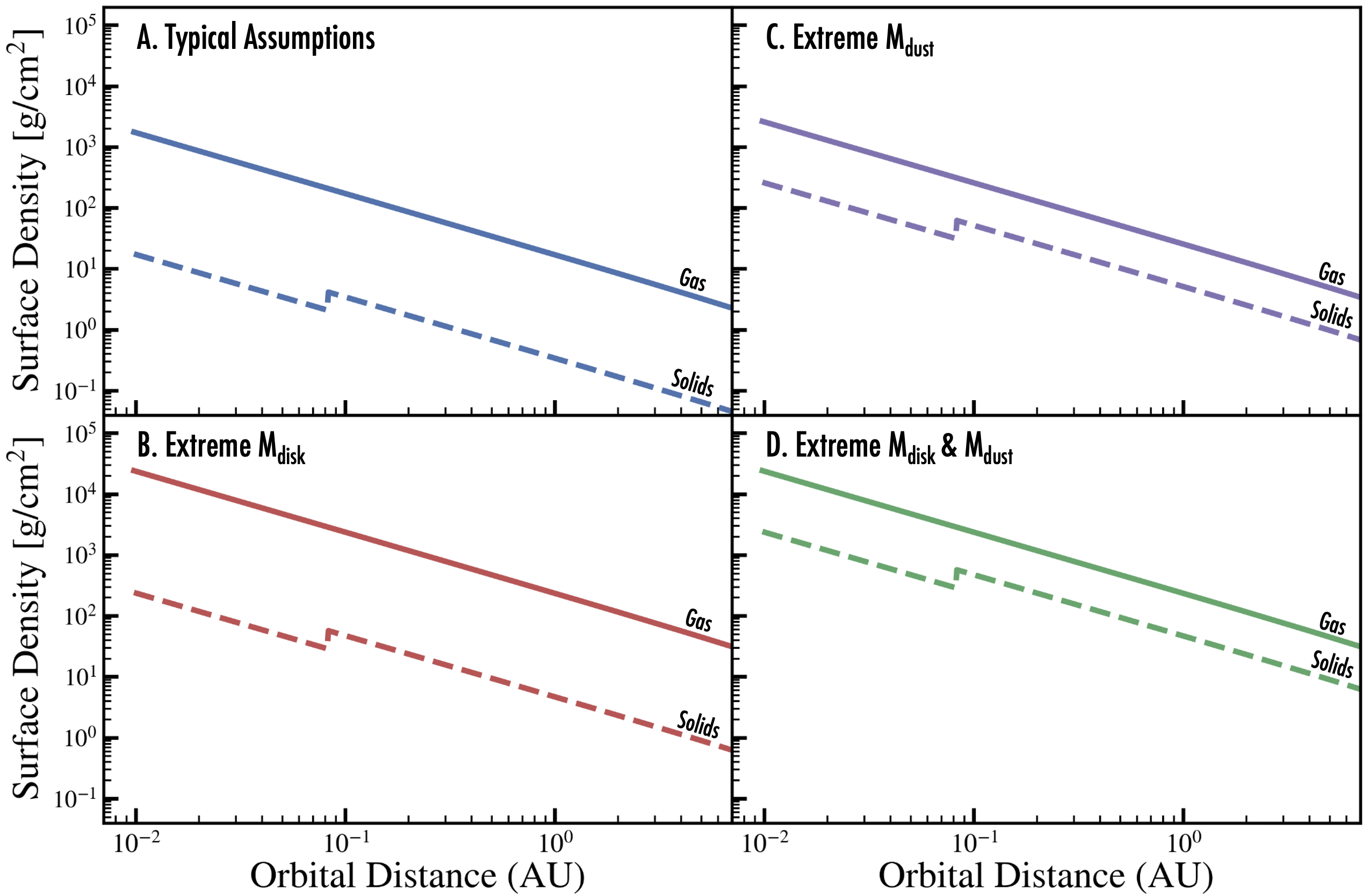}
\vspace{-0.5cm}
\end{center}
\label{fig:disks}
\end{figure}
\noindent \textbf{Fig~S6.} Surface density profiles of the initially assumed gas and dust in the disks for the four \textbf{(A-D)} different models considered.

\clearpage

\begin{table}[H]
\centering
\begin{tabular}{|c|c|}
\hline
\textbf{Barycentric Julian Date,}  &  \textbf{Radial velocity,}  \\
\textbf{BJD}                       &  [$\mathbf{\mathrm{m\:s^{-1}}}$]  \\\hline
 2458854.02373                     &  $-8.7 \pm 8.1$  \\
\hline
\end{tabular}
\end{table}
\noindent {\bf Table~S1.} \textbf{HPF radial velocity time series.} A subset of the data is shown here. A machine-readable version of the full dataset is available in Data S1.

\clearpage

\begin{table}[H]
\centering
\begin{tabular}{|c|c|c|c|c|c|}
\hline
\textbf{BJD} & \textbf{dLW,}                       &   \textbf{CRX,}                            &   \textbf{Ca II IRT 1} &  \textbf{Ca II IRT 2} &  \textbf{Ca II IRT 3} \\ 
             & [$\mathbf{\mathrm{m^{2}\:s^{-2}}}$] &   [$\mathbf{\mathrm{m\:s^{-1}\:Np^{-1}}}$] &                        &                       &   \\ \hline
 2458854.02373 &   $69.5 \pm 18.9$ &   $-205.6 \pm 86.9$ &  $0.550 \pm 0.003$ &  $0.308 \pm 0.002$ &  $0.329 \pm 0.002$ \\
\hline
\end{tabular}
\end{table}

\noindent \textbf{Table S2. HPF activity indicator time series.} The columns are: Barycentric Julian Date (BJD), Differential Line Width (dLW), Chromatic Index (CRX), and the Calcium II Infrared Triplet (IRT) index for the three separate Ca II IRT lines (Ca II IRT 1, 2 and 3). A subset of the data is shown here. A machine-readable version of the full dataset is available in Data S2.

\clearpage

\begin{table}
\centering
\begin{tabular}{|c|c|c|c|}
\hline
\multicolumn{4}{|c|}{\textbf{Fit Parameters}}           \\ \hline
\textbf{Parameter}              &                             \textbf{Description }  &   \textbf{Prior}                      & \textbf{Posterior}                \\\hline
                            $P$ &                              Orbital period (days) &   $\mathcal{U}$(3,4)                  &  $3.71778_{-0.00081}^{+0.00080}$  \\\hline
 $T_{\mathrm{conj}}$-2,458,800  &       Time of Conjunction $(\mathrm{BJD_{TDB}})$   &   $\mathcal{U}$(72,76)                &  $74.26_{-0.14}^{+0.14}$          \\\hline
                            $e$ &                                       Eccentricity &   $\mathcal{U}$(0,0.9)                &  $0.076_{-0.047}^{+0.057}$        \\\hline
                       $\omega$ &                  Argument of periastron ($^\circ$) &   $\mathcal{U}$(0,360)                &  $82_{-47}^{+102}$                \\\hline
                            $K$ &                            RV semi-amplitude (m/s) &   $\mathcal{U}$(0,100)                &  $23.4_{-1.4}^{+1.5}$             \\\hline
        $\sigma_{\mathrm{HPF}}$ &                                HPF RV jitter (m/s) &   $\mathcal{J}$(0.1,200)              &  $1.17_{-0.91}^{+2.2}$            \\\hline
                       $\gamma$ &                                HPF RV offset (m/s) &   $\mathcal{U}$(-100,100)             &  $-2.2_{-1.1}^{+1.1}$             \\\hline
\multicolumn{4}{|c|}{\textbf{Derived Parameters}}           \\ \hline
 $T_{\mathrm{peri}}$-2,458,800  &        Time of Periastron $(\mathrm{BJD_{TDB}})$   &                                   -   &  $74.02_{-0.57}^{+0.67}$          \\\hline
                     $m \sin i$ &                           Planet mass ($M_\oplus$) &                                   -   &  $13.15_{-0.82}^{+0.84}$          \\\hline
                            $a$ &                       Orbital semimajor axis (au)  &                                   -   &  $0.02262\pm0.00018$              \\\hline
\end{tabular}
\end{table}
\noindent \textbf{Table S3. Summary of priors and resulting parameters from the HPF RV fit}. $\mathcal{N}(m,\sigma)$ denotes a normal prior with mean $m$, and standard deviation $\sigma$; $\mathcal{U}(a,b)$ denotes a uniform prior with a start value $a$ and end value $b$, $\mathcal{J}(a,b)$ denotes a Jeffreys prior with a start value $a$ and end value $b$. The posteriors report the median values and the 68\% credible intervals.

\end{document}